\documentclass[12pt,a4paper]{article}

\usepackage{amsmath,amssymb}
\usepackage[bookmarks=true]{hyperref}
\usepackage[nosort]{cite}
\usepackage[bulletsep]{collect}
\def\gfxon{\usepackage[final]{graphicx}}

\gfxon

\usepackage[T1]{fontenc}
\usepackage{amsthm}
\usepackage{amsfonts}
\usepackage{bm}
\usepackage{cancel}

\makeatletter
\let\old@startsection=\@startsection
\renewcommand{\@startsection}[6]{\old@startsection{#1}{#2}{#3}{#4}{#5}{#6\mathversion{bold}}}
\makeatother

\def\eeq{\end{eqnarray}}

\def\p{\partial}

\def\=:{=\hspace{-.7em}\raisebox{1.1ex}{.}\hspace{.1em}\raisebox{-0.2ex}{.} }

\newcommand{\beqn}{\begin{eqnarray}}
\newcommand{\eeqn}{\end{eqnarray}}

\newcommand {\beq}{\begin{eqnarray}}
\newcommand {\eeqq}{\end{eqnarray}}

\newcommand {\Tr}{{\rm Tr}\,}

\makeatletter \@addtoreset{equation}{section} \makeatother

\usepackage[width=0.90\textwidth,font={small},labelfont={bf,small}]{caption}

\makeatletter
\def\mr@ignsp#1 {\ifx\:#1\@empty\else #1\expandafter\mr@ignsp\fi}%
\newcommand{\multiref}[1]{\begingroup
\xdef\mr@no@sparg{\expandafter\mr@ignsp#1 \: }%
\def\mr@comma{}%
\@for\mr@refs:=\mr@no@sparg\do{\mr@comma\def\mr@comma{,}\ref{\mr@refs}}%
\endgroup}
\makeatother

\ifx\href\asklfhas\newcommand{\href}[2]{#2}\fi

\begin{document}

\begin{flushright}\footnotesize
\texttt{IFUP-TH/2012-11} \\
\texttt{}
\vspace{0.5cm}
\end{flushright}
\vspace{0.3cm}
\renewcommand{\thefootnote}{\arabic{footnote}}
\setcounter{footnote}{0}
\begin{center}%
{\Large\textbf{\mathversion{bold}
Spontaneous Magnetization through Non-Abelian Vortex Formation in Rotating Dense Quark Matter
}
\par}

\vspace{1cm}%

\textsc{Walter Vinci$^{1}$, Mattia Cipriani$^{1}$ and Muneto Nitta$^{2}$}

\vspace{10mm}
$^1$\textit{University of Pisa, Department of Physics ``E. Fermi'', INFN,\\%
Largo Bruno Pontecorvo 7, 56127, Italy}
\vspace{.3cm}
\\
$^2$\textit{Department of Physics, and Research and Education Center 
for Natural Sciences,\\ Keio University, 4-1-1 Hiyoshi, Yokohama, 
Kanagawa 223-8521, Japan}

\vspace{7mm}

\thispagestyle{empty}

\texttt{walter.vinci(at)pi.infn.it}\\ 
\texttt{cipriani(at)df.unipi.it}\\ 
\texttt{nitta(at)phys-h.keio.ac.jp}\\

\par\vspace{1cm}

\vfill

\textbf{Abstract}\vspace{5mm}

\begin{minipage}{12.7cm}
When a color superconductor of high density QCD is rotating, superfluid vortices are inevitably created along the rotation axis. 
In the color-flavor locked phase realized at the asymptotically large chemical potential, there appear non-Abelian vortices carrying both circulations of superfluid and color magnetic fluxes. 
A family of solutions has a degeneracy characterized by the Nambu-Goldstone modes ${\mathbb C}P^2$, associated with the color-flavor locked symmetry 
spontaneously broken in the vicinity of the vortex.
In this paper,  
we study electromagnetic coupling of the non-Abelian vortices and 
find that the degeneracy is removed with the induced effective potential.  
We obtain one stable vortex solution and a family of metastable vortex solutions, both of which carry ordinary magnetic fluxes in addition to color magnetic fluxes. We discuss quantum mechanical decay of the metastable vortices by quantum tunneling, and compare the effective potential with the other known potentials, 
the quantum mechanically induced potential 
and the potential induced by the strange quark mass.

\end{minipage}

\vspace{3mm}

\vspace*{\fill}

\end{center}

\newpage

\section{Introduction}

The color superconducting phase occupies the low temperature-high density region of the QCD phase diagram~\cite{Rajagopal:2000wf}. For these values of physical parameters, quarks condense in couples, similarly to what happens in ordinary superconductors for electrons forming Cooper pairs. Moreover, this phase is classified with different names, depending on which flavors participate to condensation. The first rough distinction is made between the so-called 2SC and color-flavor locked (CFL) phase~\cite{Alford:2007xm,Alford:1998mk}: 
the first one is characterized by the condensation of the two lighter flavors only; 
in the other all three flavors participate to the condensation. 
It has been discussed that the recently found heavy neutron star~\cite{Demorest:2010bx} implies the equation of state soft, which denies the existence of exotic matters including quark matters. However, there is also an objection to that argument~\cite{Masuda:2012kf,Weissenborn:2011qu} and the situation is not conclusive yet.
The CFL phase may exsist in the core of compact stars, and 
it is important to suggest some experimental signatures of its exsistence.

In the CFL phase, due to the condensation of the three quark flavors, the original symmetry group is broken down as $G= U(1)_{B} \times SU(3)_{C} \times SU(3)_{F} \rightarrow H = SU(3)_{C+F} \times \mathbb{Z}_{3}$, where $U(1)_{B}$ is the baryon symmetry group, while, apart from the color and flavor symmetries, $SU(3)_{C+F}$ is the color-flavor locked group. 
Since $U(1)_{B}$ and $SU(3)_{C}$ are spontaneously broken in the ground state, 
it is superfluid and color superconducting, respectively, at the same time.
When a color superconductor is rotating, which will be the case if it is realized in a neutron star core, superfluid $U(1)_{B}$ vortices~\cite{Forbes:2001gj,Iida:2002ev,Iida:2004if} are inevitably created along the rotation axis, 
as in usual superfluids such as $^4$He and Bose-Einstein condensates in ultracold atomic gases. 
However a $U(1)$ vortex is unstable to decay into 
a set of three non-Abelian semi-superfluid vortices 
found by Balachandran, Digal and Matsuura~\cite{Balachandran:2005ev},\footnote{Non-Abelian vortices were found in the context of supersymmetric theories~\cite{Auzzi:2003fs,Hanany:2003hp}, in models whose vacuum possesses the color-flavor locking structure as the CFL does. 
See Refs.~\cite{Tong:2005un,Eto:2006pg,Shifman:2007ce,Tong:2008qd} 
as a review in this context.
} 
because of a long range repulsion between them~\cite{Nakano:2007dr}. 
Then such non-Abelian vortices are expected to form a vortex lattice~\cite{Nakano:2007dr,Nakano:2008dc}.
The precise numerical solutions and analytic expression of the asymptotic form of the vortex solution were given in~\cite{Eto:2009kg}.
Since a non-Abelian vortex winds $2\pi/3$ of the $U(1)_{B}$ phase and 
$SU(3)_{C}$ at the same time, it carries 1/3 circulation of that 
of the $U(1)_{B}$ vortex and a color magnetic flux, respectively. 
Non-Abelian vortices are superfluid vortices as well as 
color magnetic flux tubes 
so that they are the most fundamental vortices 
in color superconductors.


These solutions are invariant only under the action of a subgroup of the color-flavor symmetry of the ground state, thus producing the further breaking $SU(3)_{C+F} \rightarrow SU(2)_{C+F} \times U(1)_{C+F}$. Consequently, the Nambu-Goldstone modes, called {\it orientational moduli},  
\begin{equation}
	\mathcal{M} = \frac{SU(3)_{C+F}}{SU(2)_{C+F} \times U(1)_{C+F}} \simeq \mathbb{C}P^{2}  \label{eq:zeromodes}
\end{equation}
arise from this breaking near the vortex core~\cite{Nakano:2007dr}. 
There is a whole family of solutions all having the same tension which can be mapped the ones into the others by color-flavor transformations.
It was shown in~\cite{Eto:2009bh} that the orientational zero modes~(\ref{eq:zeromodes}) are in fact normalizable, and the effective world-sheet theory was constructed as a ${\mathbb C}P^2$ non-linear sigma model. 
The effect of the strange quark mass was taken into account in 
the effective world-sheet theory as the potential term~\cite{Eto:2009tr}.
In the asymptotically large chemical potential where strange quark mass can be neglected, quantum effects imply the presence of monopoles, seen as kinks on the vortex worldsheet~\cite{Eto:2011mk,Gorsky:2011hd}.
Coupling of non-Abelian vortices to quasi-particles such as gluons, $U(1)_{\rm B}$ Nambu-Goldstone bosons (phonons) were determined~\cite{Hirono:2010gq}.  
More recently, the electro-magnetic coupling to
the non-Abelian vortices has been studied and 
it has been found that a lattice of non-Abelian vortices 
acts as a polarizer of photons~\cite{Hirono:2012ki}. 
Some implications of the existence of non-Abelian vortices in neutron stars 
were studied~\cite{Sedrakian:2008ay,Shahabasyan:2009zz,Shahabasyan:2011aa}.
These analysis were all based on the Ginzburg-Landau model which is valid for temperatures near the critical one~\cite{Giannakis:2001wz,Iida:2000ha,Abuki:2006dv}. 
In addition to the analyses based on the Ginzburg-Landau model, 
the Bogoliubov-de Gennes approach which is valid below the critical temperature was also used to study the fermionic structures of the non-Abelian vortices, 
and it was found that non-Abelian vortices trap Majorana fermion zero modes belonging to the triplet representation of the unbroken symmetry $SU(2)_{\rm C+F}$ at their cores~\cite{Yasui:2010yw,Fujiwara:2011za}. 
One of interesting consequences of the fermion zero modes is that the exchange statistics of non-Abelian vortices makes them non-Abelian anyons~\cite{Yasui:2010yh}.

In this paper, we study electromagnetic coupling of non-Abelian vortices based on the Ginzburg-Landau model. 
While the analysis in~\cite{Hirono:2012ki} neglected the mixing of the photon and the gluon, we correctly taking it into account.
Our analysis is, in some sense, a continuation of the work done in~\cite{Balachandran:2005ev}, where also electromagnetic coupling was considered. 
Since the $U(1)_{\textsc {em}}$ symmetry is embedded into the flavor $SU(3)_{\rm F}$ symmetry, the latter symmetry is not intact anymore;  
only its subgroup $SU(2)_{\rm F} \times U(1)_{\rm F}$ remains exact. 
We find, as a consequence, that it removes the degeneracy of solutions~(\ref{eq:zeromodes}) in energy between the different vortex solutions.
We show that the solution of~\cite{Balachandran:2005ev}, which we call ``the Balachandran-Digal-Matsuura (BDM) vortex'', represents the lowest energy configuration when we neglect the strange quark mass, which was in fact stated in~\cite{Balachandran:2005ev}. We are able also to identify two other different ``diagonal'' vortex solutions having a tension slightly higher compared to the former, and being related together by the residual $SU(2)_{C+F}$ and then constituting a full family of solutions, a $\mathbb{C}P^{1}$ space as a submanifold of the $\mathbb{C}P^{2}$ moduli space. We refer to these last solutions as ``$\mathbb{C}P^{1}$ vortices''. We find another special vortex solution whose winding does not regard the direction in the color space that allows for the electromagnetic gauge field to couple, namely the $T^{8}$ direction in $SU(3)_{C}$. This solution possesses the highest tension and should be unstable. We call it  ``pure color'' vortex. An energy potential connects all the different vortices: the BDM solution is its absolute minimum, while the two $\mathbb{C}P^{1}$ vortices are local minima, along with their full $\mathbb{C}P^{1}$ space; the pure color vortex stays at the maximum. 
We then find that the BDM and $\mathbb{C}P^{1}$ vortices are 
spontaneously magnetized to carry magnetic fluxes of the electromagnetic field, in addition to the color magnetic fluxes. 
The magnetic flux of the BDM vortex is $-2$ times the one of the $\mathbb{C}P^{1}$ vortices. 
Since the minimal color-less bound state is obtained as the $U(1)_{\rm B}$ vortex made of one BDM vortex and two $\mathbb CP^{1}$ vortices, this color-neutral bound state of vortices necessarily carries also no electromagnetic flux, and vice versa. The amount of magnetic flux is proportional to $e^2/g^2$ and is tiny.

This situation is changed when the strange quark mass is taken into account. In this case, as discussed in Ref.~\cite{Eto:2009tr}, a potential arises between the vortex configurations considered so far, breaking the residual $\mathbb{C}P^{1}$ symmetry and lowering the tension of one of the $\mathbb{C}P^{1}$ vortices, which will be called $\mathbb{C}P^{1}_{+}$ vortex. This solution has now the lowest energy and is then the most stable one, to which the BDM vortex can decay. We will show that, for densities typical of the inner neutron star core, the $\mathbb{C}P^{1}_{+}$ vortex is the relevant solution, instead of the previous considered BDM vortex.

This paper is organized as follows. In Section~\ref{sec:CFLphase} we introduce the theoretical setting in the case without electromagnetic coupling and we review the pure color solutions. In Section~\ref{sec:EMcoupling} we add the electromagnetic coupling to our model and we explore the different features of the three kinds of vortices mentioned above: the BDM vortex, the $\mathbb{C}P^{1}$ vortices and the pure color vortex. We calculate magnetic fluxes which these vortices carry. We then discuss the $\mathbb{C}P^{1}$ vortices, which are metastable classically, decay quantum mechanically into the BDM vortex by a quantum tunneling. We give an estimate of the decay probability.
Then in Section~\ref{sec:stability} we compare the potential induced by the electromagnetic interaction which we found with other potential terms; 
the one quantum mechanically induced and the one induced by a non-zero strange quark mass. 
In Section~\ref{sec:Conclusions} we present our conclusions. 
In Appendix, we summarize equations for non-Abelian vortices.

\section{QCD in the CFL Phase}\label{sec:CFLphase}

\subsection{Symmetry breaking in the CFL hase}
Color-superconductivity in hight density QCD is almost ``inevitable''. When the chemical potential $\mu$ is large enough, $\mu>\Lambda_{QCD}$, QCD is in a perturbative regime. In this regime, it is easy to see that gauge interactions mediate attractive forces between quarks. The BCS mechanism of superconductivity implies then formation of Cooper pairs and of  a diquark condensate~\cite{Alford:2007xm}. In the most symmetric phase, (color-flavor-locked, CFL), the diquark condensate has the following form~\cite{Alford:1998mk}:   
\begin{eqnarray}
\left<\psi_{i}^{\alpha}C\gamma_{5}\psi_{j}^{\beta}\right>& = & \epsilon^{\alpha \beta \gamma}\epsilon_{ijk}\Phi_{\gamma}^{\ k} , \quad i,j,k=u,d,s\quad \alpha,\beta,\gamma=r,g,b\,,
\label{eq:condensate}
\end{eqnarray}
where the two quarks pair in a parity-even, spin singlet channel, while the color and flavor wave functions are completely antisymmetric. The order parameter $\Phi$ is a 3 by 3 matrix that can be written, in terms of pair condensation, as follows
\begin{eqnarray}
\Phi_{\gamma}^{\ k}=  \left(
\begin{array}{ccc}
 \left< ds \right>_{gb} &  \left< us \right>_{gb} &  \left< ud \right>_{gb} \\
 \left< ds \right>_{rb} &  \left< us \right>_{rb} &  \left< ud \right>_{rb} \\
 \left< ds \right>_{rg} &  \left< us \right>_{rg} &  \left< ud \right>_{rg}  
\end{array}
\right)\,.
\label{eq:comb}
\end{eqnarray}
It transforms under gauge and flavor symmetries as a $(3_{C},\bar 3_{F})$ representation :
\begin{equation}
\Phi \rightarrow U_{C}\Phi U_{F}^{\dagger}\, .
\label{eq:color-flavor}
\end{equation}
In the CFL phase, the ground state is given by the following value of the order parameter (up to color-flavor rotations):
\begin{equation}
\left<\Phi_{\gamma}^{\ k}\right>= \Delta_{\textsc{cfl}}\delta_{\gamma}^{\ k}\,.
\label{eq:vacc}
\end{equation}

We notice \emph{en passant} that other phases are possible when one considers a non-zero mass for the strange quark. For example, when the chemical potential is of order of $\mu\sim m_{s}^{2}/\Delta_{\textsc{cfl}}$, the strange quark does not participate to condensation anymore and we have the so-called 2SC phase, where the order parameter is given by $\Phi_{\gamma}^{\ k}=\Delta_{2\textsc{sc}}\delta_{\gamma}^{\ 3}\delta_{3}^{\ k}$.

As well-known,  QCD has the following symmetries when quark masses are neglected:
\begin{equation}
G=U(1)_{B}\times SU(3)_{C}\times SU(3)_{L}\times SU(3)_{R}\, .
\label{eq:symmetry}
\end{equation}
The condensate in Eq.~(\ref{eq:vacc}) breaks them to the following residual group:
\begin{equation}
H=SU(3)_{C+L+R}\, .
\end{equation}
The residual non-Abelian symmetry in the ground state is a peculiarity of the CFL ground state and the origin of some of the most interesting properties of this phase. Notice for example that chiral symmetry is broken perturbatively.

The symmetry breaking pattern described above implies the existence of stable topological ``semi-superfluid'' vortices~\cite{Balachandran:2005ev}
\begin{equation}
\pi_{1}(G/H)=\pi_{1}\left(\frac{U(1)_{B}\times SU(3)_{C-F}}{\mathbb Z_{3}}\right) \simeq \mathbb Z\,.
\label{eq:top}
\end{equation}
The origin of the word semi-superfluid arises because the smallest non-trivial loops  corresponding to vortices involve both global and gauge rotations. Vortices in the CFL phase of QCD are similar to vortices appearing in both superfluid and superconductors.

In the next Section we will briefly review the construction and the properties of these vortices.

\subsection{Landau-Ginzburg description of the CFL phase}
The Landau-Ginzburg description of the CFL phase in terms of a field theory for the order parameter is appropriate at temperatures close to the critical temperature $T_{c}$ for the CFL phase transition. Since we will always be interested in static configurations, we write down only the terms that do not include time derivatives~\cite{Giannakis:2001wz,Iida:2000ha,Iida:2001pg}:
\begin{eqnarray}
\mathcal L_{LG} & = &  \Tr\left[\frac{\epsilon \lambda}{2}F_{0j}F^{0j}-\frac1{4}F_{ij}F^{ij}+ K_{1}\nabla_{i} \Phi^{\dagger}\nabla^{i} \Phi -\lambda_{2}(\Phi^{\dagger}\Phi)^{2}+m^{2}\Phi^{\dagger}\Phi  \right] + \nonumber \\
& -& \lambda_{1}(\Tr[\Phi^{\dagger}\Phi])^{2}-\frac{3 m^{4}}{4(3\lambda_{1}+\lambda_{2})}\, .
  \label{eq:lagrangianNoEM}
\end{eqnarray}
Traces are taken on both color and flavor indexes when as appropriate. We use the following conventions:
\begin{eqnarray}
& F_{ij}= \p_{i}A_{j}-\p_{j}A_{i}-ig_{s}[A_{i},A_{j}],& \nonumber \\
&  \nabla_{i}\Phi=\partial_{i}\Phi-ig_{s} A_{i}\Phi,& \nonumber \\
& A_{i}=A^{a}_{i}T^{a}, \quad F_{ij}\equiv F_{ij}^{a}T^{a}, \quad     \Tr(T^{a}T^{b})=\delta^{ab}.   &
\label{eq:conv}
\end{eqnarray}
The coefficient in the expression above may be calculated directly from the QCD lagrangian using perturbative techniques. We quote here the standard results obtained in literature through perturbative calculations in QCD~\cite{Iida:2002ev,Giannakis:2001wz,Iida:2000ha}:
\begin{align}
&2 \lambda_{1}=2 \lambda_{2}=3K_{1} =  \frac{7 \zeta(3)}{4 (\pi T_{c})^{2}} N(\mu) \, , \nonumber \\
&m^{2}=-4 N(\mu) \log \frac{T}{T_{c}},\quad N(\mu)=\frac{\mu^{2}}{2\pi^{2}} \, ,\nonumber \\
&g_{s}=\sqrt\frac{24 \pi^{2}\lambda}{27 \log \mu/\Lambda}\,,\quad T_{c}\sim\mu \exp\left(-\frac{3 \pi^{2}}{\sqrt2 g_{s}}\right) \, ,
\end{align}
where $\mu$ is the chemical potential, $\Lambda$ the QCD scale and $T_{c}$ the critical temperature, while $\epsilon$ and $\lambda$ are the dielectric and diamagnetic constants. In the following, we assume electric fields to be vanishing, and we thus neglect the first term in Eq.~(\ref{eq:lagrangianNoEM}).

The expectation value of $\Phi$ can be directly found by minimizing the potential of expression~(\ref{eq:lagrangianNoEM}), obtaining:
\begin{eqnarray}
\left<\Phi\right>=\Delta_{\textsc{cfl}} {\bf 1}_{3},\quad \Delta_{\textsc{cfl}}^{2}\equiv\frac{m^{2}}{2(3\lambda_{1}+\lambda_{2})}\,.
\label{eq:vacuumLG}
\end{eqnarray}
Masses of gauge bosons and scalars are given by the following:
\begin{eqnarray}
m^{2}_{g}=2 g^{2}_{s}\Delta_{\textsc{cfl}}^{2}K_{1}, \quad m^{2}_{\phi}=\frac{2 m^{2}}{K_{1}},\quad m_{\chi}^{2}=\frac{4 \lambda_{2}\Delta_{\textsc{cfl}}^{2}}{K_{1}},\quad m_{\varphi}^{2}=0,
\end{eqnarray}
where $\varphi$ is the massless Nambu-Goldstone boson related to the breaking of $U(1)_{B}$ symmetry, and $\phi$ and $\chi$ are respectively the trace and traceless part of $\Phi$.

Equations of motion can also be directly found from the lagrangian above, and they read as:
\begin{eqnarray}
\nabla_{i}F^{ij} & = & -i g_{s}K_{1} \left[  \nabla_{j}\Phi \Phi^{\dagger}-\Phi(\nabla_{j}\Phi)^{\dagger}-\frac1N \Tr\left(\nabla_{j}\Phi \Phi^{\dagger}-\Phi(\nabla_{j}\Phi)^{\dagger}\right)\right]\,, \nonumber \\
\nabla_{j}\nabla^{j}\Phi &=& \frac{2}{K_{1}} \left[-\lambda_{2} \Phi \Phi^{\dagger} -2 \lambda_{1}\Tr(\Phi^{\dagger}\Phi)+m^{2}\right] \Phi\,.
\label{eq:motion}
\end{eqnarray}

\subsection{``Pure'' color-magnetic flux tubes}
Let us discuss now in more detail the color flux tubes that are stabilized by  the topology of the CFL phase. These solution have been thoroughly studied in literature~\cite{Balachandran:2005ev,Eto:2009bh,Nakano:2007dr,Eto:2009kg} and have been dubbed, in various works, as semi-superfluid, color-magnetic or non-Abelian strings. { 
In the following we will also refer to these vortices as {\itshape un-coupled} vortices, because in Section~\ref{sec:EMcoupling} we will couple them to the electromagnetic gauge field. The latter solutions will obviously be the {\itshape coupled} ones.}

Strings in the CFL are semi-superfluid because the fundamental vortex corresponds to a non-trivial closed loop constructed with both global $U(1)_{B}$ and gauge $SU(3)_{C}$ symmetry. This fact can be easily seen by more closely inspecting the relevant  homotopy group in Eq.~(\ref{eq:top}). The $\mathbb Z_{3}$ factor in the denominator implies that the smallest loop can be constructed by picking up a $\pi/3$ global phase and an additional $2\pi/3$ gauge phase. More concretely, a vortex configuration can be written as follows: 
\begin{equation}
\Phi(r,\varphi)_{\bar r}= \Delta_{\textsc{cfl}}
\left(
\begin{array}{ccc}
 e^{i\varphi}f(r) &  0 & 0  \\
0  &  g(r) &  0 \\
 0 & 0  & g(r)   
\end{array}
\right),\quad A_{i}^{8}T^{8}=\sqrt\frac23 \frac1g_{s}\frac{\epsilon_{ij} x^{j}}{r^{2}}[1-h(r)] \, T^{8} \,,
\label{eq:antiredvortex}
\end{equation}
where the profile functions $f(r)$, $g(r)$ and $h(r)$ satisfy the equations reported in the Appendix and satisfy the following boundary conditions:
\begin{equation}
f(0)=0,\quad g'(0)=0,\quad h(0)=1,\quad f(\infty)=g(\infty)=1,\quad h(\infty)=0\,.
\end{equation}
The vortex above corresponds to the following closed loop in the global-gauge group:
\begin{equation}
\Phi(\infty,\varphi)=e^{i\varphi/3}e^{-i\sqrt{{2}/{3}} \, \varphi \,  T^{8}}\Phi(\infty,0)\,.
\label{eq:winding}
\end{equation}
Since the $\left<ds\right>_{gb}$ pairing winds in this case, we called the previous configuration ``anti-red'' vortex.

 The tension of a global vortex can be conveniently written as a sum of a divergent term, typical of global vortices, plus a finite term.
\begin{equation}
\mathcal T=\mathcal T_{div}+\mathcal T_{fin}.
\end{equation}
By inserting the asymptotical value above into the original lagrangian~(\ref{eq:lagrangianNoEM}), we can extract the logarithmically divergent contribution:
\begin{align}
 \mathcal T_{div}&= 2 \pi K_{1}\int {r dr \,\Tr |\nabla_{i}\Phi|^{2}}\simeq  2 \pi K_{1} \int r dr\Tr \left[\frac1r \p_{\varphi}\Phi+i \frac1r T\Phi\right] =  \nonumber \\
&= \frac{2\pi}{3}\Delta_{\textsc{cfl}}^{2}K_{1}\ln L\,, \qquad r\rightarrow \infty
\label{eq:divergent}
\end{align}
where $L$  denotes the typical size of the system. 
The divergent contribution to the  tension can be put in relation with a generic fractionally quantized circulation $c_{B}$ of the superfluid quarks:
\begin{align}
\mathcal T_{div} =& \frac{2\pi}{3}\Delta_{\textsc{cfl}}^{2}K_{1} \ln L=6\pi c_{B}^{2} \Delta_{\textsc{cfl}}^{2}K_{1} \ln L\,,\nonumber \\
c_{B}\equiv & \frac{m^{*}}{2 \pi}\oint \vec v_{sup}\cdot d\vec l  =\frac{n}{3}\,.
\end{align}
 
 Semi-superfluid vortices in the CFL phase are also color-magnetic in the sense that they also carry a non-trivial color flux. For the configuration above we have:
 \begin{align}
& \Phi^{8}_{flux}= \oint \vec A^{8}\cdot  d\vec l=\sqrt\frac{2}{3}\frac{2 \pi}{g_{s}}\,.
 \end{align}

\begin{figure}[htbp]
\begin{center}
\includegraphics{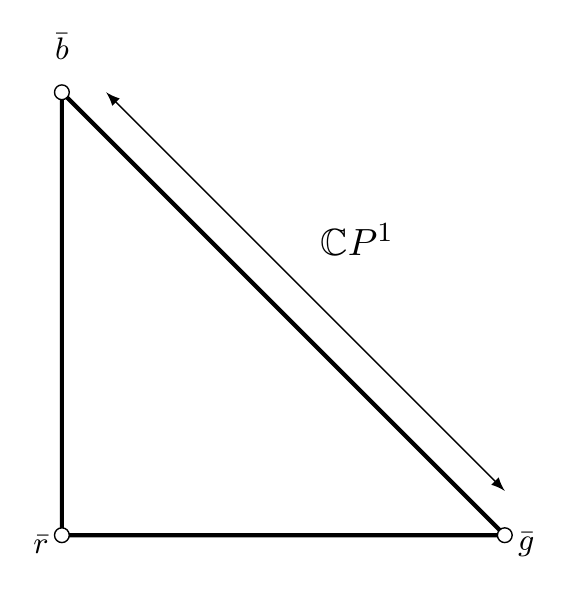}
\caption{The ``toric diagram'' of $\mathbb{C}P^2$. The different vortices $\bar{r},\bar{g},\bar{b}$ are located at the vertices: the winding of these vortices is localized in one of the diagonal entries in the matrix field $\Phi$. The upper-right edge of the diagram corresponds to the $\mathbb{C}P^1$ subspace invariant under the $SU(2)_{C+F}$, whose generators commute with $T^{\textsc{em}}$ defined in Section~\ref{sec:EMcoupling}.}
\label{fig:Toric}
\end{center}
\end{figure}

Since the vortex configuration~(\ref{eq:antiredvortex}) breaks the $SU(3)_{C+F}$ down to a $U(1)_{C+F}$, we can straightforwardly construct the most general vortex configuration by applying color-flavor rotations:
\begin{eqnarray}
\Phi(r,\varphi)_{\text{general}}=U\Phi(r,\varphi)_{\bar r}U^{-1}\,,
\end{eqnarray}
which gives a full set of degenerate configurations given 
by the following space:
\begin{eqnarray}
\mathcal M_{\textsc{cfl}} \simeq 
\frac{SU(3)}{SU(2)\times U(1)}=\mathbb C P^{2}\,.
\end{eqnarray}
We can schematically represent the space $\mathbb C P^{2}$ in terms of its ``toric'' diagram, as in Fig.~\ref{fig:Toric}  Each point in the internal part of the diagram represents a 2-torus generated by the $U(1)^{2}$ isometry of  $\mathbb C P^{2}$. Each point on the sides represents a torus where a $U(1)$ degenerates. The vertices of the triangles represent just points in the moduli space. In our notations, the vertices are represented by the 3 diagonal vortex configurations where the winding is concentrated on each diagonal entry only, giving respectively $\bar r$, $\bar g$ and $\bar b$ vortices. 

\begin{figure}[htbp]
\begin{center}
\includegraphics[width=8cm]{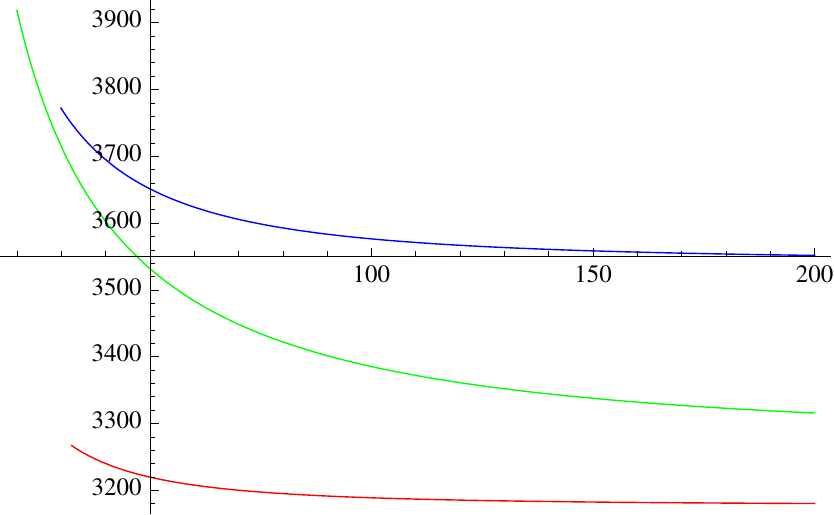}
\caption{The figure shows the monotonic decreasing of the corrections to the tension of the vortex as function of the gauge coupling, which in this figure is plot against the mass of the gauge bosons $m_{g}$. The blue line corresponds to the ``realistic'' parameters: $\mu\sim 500$MeV, $\Lambda\sim200$Mev, $T_{c}\sim 10$MeV, $T\sim 0.9 T_{c}$, which correspond to $\Delta_{\textsc{cfl}}=$7 Mev, $K_{1}=9$, $m_{\phi}=34$Mev, $m_{\chi}=17$Mev. The red and green lines correspond respectively to $(m_{\phi},m_{\chi})=(50$Mev,$10$Mev) and $(m_{\phi},m_{\chi})=(10$Mev,$50$Mev). The logarithmically divergent tension has been cut-off at a distance $L=0.4$MeV$^{-1}$. Notice that the choice of $L$ is arbitrary. A different value would correspond to a logarithmic shift in the total tension, but the monotonic decrease of the tension would be the same, since it is only given by the dependence on $g_{s}$ of the finite term $\mathcal T_{fin}.$}
\label{fig:tension}
\end{center}
\end{figure}

To conclude this Section, we discuss the dependence of the energy on the coupling constant $g_{s}$. As shown by Eq.~(\ref{eq:divergent}) the logarithmically divergent contribution to the tension does not depend on the gauge coupling. However, finite corrections to the energy do depend  in general on the gauge coupling and on the various coefficients in the potential energy. We have calculated these finite contributions numerically and we have studied their behavior under the change of $g_{s}$. Fig.~\ref{fig:tension} shows that the tension decreases monotonically with increasing of the gauge coupling. The behavior is generic and independent of the values of scalar masses. It can be intuitively understood noticing the $1/g_{s}^{2}$ dependence in the kinetic terms for the gauge potentials. 

\section{CFL Phase with Electromagnetic Interactions}\label{sec:EMcoupling}

\subsection{Electromagnetic interactions}
The introduction of electromagnetic interactions is straightforward once we look at the charge structure of the order parameter as in Eq.~(\ref{eq:comb}). The charges of each column of $\Phi$ are respectively $Q_{\textsc{em}}(\Phi)=(-2/3,1/3,1/3)$ and electromagnetic gauge transformations can be implemented by a right multiplication generated by:
\begin{eqnarray}
\Phi \rightarrow \Phi e^{i \alpha T^{\textsc{em}}},\quad T^{\textsc{em}}=\frac{1}{3}\ {\rm diag}(-2,1,1), \quad &T^{8}=\sqrt{3/2}\, T^{\textsc{em}} \,.
\end{eqnarray}
We can now generalize the Landau-Ginzburg description of the previous Section:
\begin{align}
\mathcal L & =  \Tr\left[-\frac14\frac32 F^{\textsc{em}}_{ij}F^{\textsc{em}ij}-\frac14F_{ij}F^{ij}+ K_{1} \nabla_{i} \Phi^{\dagger}\nabla^{i} \Phi -\lambda_{2}(\Phi^{\dagger}\Phi)^{2}+ m^{2}\Phi^{\dagger}\Phi  \right] +\nonumber \\
& \qquad \qquad - \lambda_{1}(\Tr[\Phi^{\dagger}\Phi])^{2}-\frac{3 m^{4}}{4(3\lambda_{1}+\lambda_{2})}\,,
  \label{eq:lagrangianEM}
\end{align}
where the coefficient in front of the kinetic term for the electromagnetic field is due to the different normalization for the $T^{\textsc{em}}$ generator; we have also used the following conventions:
\begin{eqnarray}
&F^{\textsc{em}}_{ij}= \p_{i}A^{\textsc{em}}_{j}-\p_{j}A^{\textsc{em}}_{i}, \quad A^{\textsc{em}}_{i}=A^{em}_{j}T^{\textsc{em}},& \nonumber \\
&  \nabla_{i}\Phi=\partial_{i}\Phi-ig_{s} A_{i}\Phi-i e\Phi A^{\textsc{em}}_{i}\,.& 
\end{eqnarray}

The ground state of the theory is unchanged with respect to the un-coupled case:
\begin{equation}
\left<\Phi_{\gamma}^{\ k}\right>= \Delta_{\textsc{cfl}}\delta_{\gamma}^{\ k}\,,
\label{eq:vaccgaug}
\end{equation}
however, the first effect of introducing electromagnetic interactions is the explicit breaking of the $SU(3)_{F}$ flavor symmetry, even if we still consider all quarks to be massless. Actually, introducing electromagnetic interactions is equivalent to gauging a $U(1)_{F}$ subgroup of the $SU(3)_{F}$ flavor symmetry. 
\begin{equation}
SU(3)_{F}\stackrel{T^{\textsc{em}}}{\longrightarrow }SU(2)_{F}\times U(1)_{\textsc{em}}\,.
\end{equation}
The full set of symmetries of the CFL phase of QCD with electromagnetic interactions is thus given by the following (we omit chiral symmetry):
\begin{equation}
G=U(1)_{B}\times U(1)_{\textsc{em}}\times SU(3)_{C}\times SU(2)_{F}\, .
\label{eq:symmetrygaug}
\end{equation}

The second important effect of introducing electromagnetic interactions in the CFL ground state is the presence of  an unbroken $U(1)_{0}$ symmetry which is a combination of  $T^{\textsc{em}}$ and the $T^{8}$ generator of $SU(3)_{C}$. They have to be rewritten in terms of their massless and massive combinations~\cite{Rajagopal:2000wf}
\begin{eqnarray}
A_{M}& =& \cos \zeta A^{em}+\sin \zeta A^{8}\,;\nonumber \\
A_{0}& =& -\sin \zeta A^{em}+\cos \zeta A^{8}\,,
\label{eq:diagon}
\end{eqnarray}
written in terms of a mixing angle $\zeta$
\begin{eqnarray}
\cos\zeta=\sqrt{\frac{e^{2}}{e^{2}+3 g^{2}_{s}/2}}\equiv\frac{e}{g_{M}}\,.
\end{eqnarray}
Notice that the mixing above is really meaningful only when the left action of $T^{8}$ and the right action of $T^{\textsc{em}}$ are really equivalent, which means for diagonal configurations of the order parameter $\Phi$. In this case, we can for example write the covariant derivative like:
\begin{equation}
\nabla_{i}\rightarrow \partial_{i}-i g_{M}A_{i}^{M}\,.
\end{equation}
Because of this non-trivial mixing, the masses of gluons are not equal in the CFL phase:
\begin{eqnarray}
m^{2}_{a}=2 g_{s}^{2}\Delta_{\textsc{cfl}}^{2}K_{1} ,\quad
m^{2}_{8}=2 g_{M}^{2}\Delta_{\textsc{cfl}}^{2}K_{1}\,,
\end{eqnarray}
where $a$ runs from 1 to 7 and $m_{8}$ is the mass of the gluon related to the generator $T^{8}$.

Apart from the unbroken gauge $U(1)$ symmetry, the CFL ground state  Eq.~(\ref{eq:vaccgaug}) has the following diagonal color-flavor symmetry
\begin{equation}
H_{\textsc{em}} = SU(2)_{C+F}\, .
\end{equation}
This reduced symmetry is crucial to understand the property of the moduli space of non-Abelian with electromagnetic coupling.

\subsection{``Coupled'' color-magnetic flux tubes}

We approach the study of vortices in the CFL phase when they are coupled to electromagnetic fields in the most general way, starting from their topological classification. The most general vortex is related to all the possible non-trivial loops in the vacuum manifold. In the CFL, this manifold is exclusively generated  by symmetry transformations:
\begin{equation}
\pi_{1}(\mathcal M_{vac})=\pi_{1}\left(G/H   \right)=\pi_{1}\left(\frac{U(1)_{B} \times SU(3)_{C-F}}{(\mathbb Z_{3})_{C-F+B}}  \right) \simeq \mathbb Z\,.
\end{equation}
The $\mathbb Z_{3}$ factor above is crucial to have semi-superfluid vortices with non-Abelian fluxes. This is due to the fact that the smallest non-trivial loop has to wind in both the global baryonic and gauge symmetry group. The conclusion above is not changed if we consider the additional electromagnetic coupling. This is due to the fact that we  can always unwind an electromagnetic phase by using the unbroken gauge group $U(1)_{0}$. However, in the un-coupled case we have a full $SU(3)$ isometry that we can use to generate all the possible vortex configurations starting with an arbitrarily chosen one, for example the one in Eq.~(\ref{eq:antiredvortex}). On the other hand, in the coupled case we only have a residual $SU(2)$ isometry, which is not enough to exhaust all possible solutions. For example, the vortex in Eq.~(\ref{eq:antiredvortex}) is invariant under this symmetry. In the toric representation of Fig.~\ref{fig:Toric} the $SU(2)$ isometry transforms points of the triangle along lines parallel to the long diagonal side. See also Fig.~\ref{fig:Toric}.

Let us make this discussion more concrete. 
A closed loop in the gauge/baryon group is given by the following transformation on the order parameter, where $\varphi$ is the angle coordinate of the space: 
\begin{eqnarray}
&\left<\Phi(\infty,\varphi)\right> 
= e^{i \theta(\varphi)} e^{i \gamma^{a}(\varphi)T^{a}}\left<\Phi(\infty,0)\right>  \,  e^{i \alpha(\varphi) T^{\textsc{em}}}  &\nonumber \\
& \Downarrow  \varphi = 2\pi& \nonumber \\
 & e^{i \gamma^{a}(2 \pi)T^{a}}  =e^{-i \theta(2 \pi)} e^{-i \alpha(2 \pi) T^{\textsc{em}}}\,,&
 \label{eq:loops}
\end{eqnarray}
where $\theta$, $\gamma^{a}$ and $\alpha$ are monotonically increasing functions.
The second line is an equation for the possible symmetry transformations giving a closed loop, or, equivalently, a possible vortex configuration. Notice that existence and stability of vortices related to various solutions of the equation above can in principle be only inferred by a direct study of equations of motion. It is possible to determine all the solutions of Eq.~(\ref{eq:loops}), for example using an explicit parameterization of the elements of $SU(3)$. In what follows we will analyze three types of solutions that cannot be related by $SU(2)$ color-flavor transformations: {1)  ``BDM'' case}, {2) ``$\mathbb CP^{1}$'' case}, and {3)  ``Pure color'' case}:

{1) \bf ``BDM'' case}

The first possibility is a closed loop generated by $T^{8}$ in  $SU(3)$ and the electromagnetic $T^{\textsc{em}}$ alone:
\begin{eqnarray}
& e^{i \gamma^{8}(\varphi) T^{8}} =e^{-i \theta(\varphi)} e^{-i \,\alpha(\varphi)T^{\textsc{em}}}&  \nonumber \\
& \Downarrow  \varphi = 2\pi& \nonumber \\
& \gamma^{8}(2\pi)/\sqrt6+\alpha(2\pi)/3=-2\pi/3, \quad \theta(2\pi)=2\pi/3\,.&
\end{eqnarray}
 The equation above determines the  phases $\gamma^{8}$ and $\alpha$ only up to a linear combination. This is a consequence of the fact that the two generators $T^{8}$ and $T^{\textsc{em}}$ are proportional and indistinguishable from each other on diagonal configurations. The configuration above is invariant under  $SU(2)$ color-flavor transformations.
%
 
{2) \bf ``$\mathbb CP^{1}$'' case}

The second possibility is a closed loop generated by winding in the $SU(3)$ group around the $T^{3}$ direction too, in addition to $T^{8}$ and $T^{\textsc{em}}$.
\begin{eqnarray}
& e^{i (\gamma^{3}(\varphi) T^{3}+\gamma^{8}(\varphi) T^{8})} =e^{-i \theta(\varphi)} e^{-i \,\alpha(\varphi)T^{\textsc{em}}}&  \nonumber \\
& \Downarrow  \varphi = 2\pi& \nonumber \\
& \gamma^{8}(2\pi)/\sqrt6+\alpha(2\pi)/3=\pi/3, \quad \gamma^{3}(2\pi)=\pm {\sqrt2} \pi,   \quad \theta(2\pi)=\pi/3.&
\end{eqnarray}
The configuration above will not be preserved by color-flavor transformations, and  we can generate a orientational moduli space using  $SU(2)$ color-flavor transformations. In fact, we have the most general configuration of this type in the form:
\begin{eqnarray}\label{eq:CFModuli}
\gamma^{3}(2\pi)T^{3}\rightarrow \gamma^{b}(2\pi)T^{b}\,\quad |\gamma^{b}|= {\sqrt2} \pi, \quad b=1,2,3\,,
\end{eqnarray}
where $\rightarrow$ denotes a replacement; the $T^{a}$ above are the generators of $SU(3)$ that commute with $T^{8}$ and form an $SU(2)$ subgroup.

{3) \bf ``Pure color'' case}

In terms of the vector $ |\gamma^{a}|$ introduced above, the two previous cases correspond to $ |\gamma^{a}|=0$ and $ |\gamma^{a}|={\sqrt2}\pi$. These are the only two cases for which a non-trivial electromagnetic phase is allowed. In all the other cases, we must have:

\begin{eqnarray}
&e^{i \gamma^{a}(\varphi)T^{a}}=e^{-i\theta(\varphi)};& \nonumber \\
& \Downarrow  \varphi = 2\pi& \nonumber \\
&\theta(2\pi)=2 \pi/3,\quad \alpha(2\pi)=0\,.&
\end{eqnarray}
We called this case ``pure color'' because it doesn't involve  electromagnetic transformations, and is thus equivalent to the well-known case of pure color vortices without the electromagnetic coupling. Notice that this case spans a whole $\mathbb CP^{2}$, while $SU(2)$ color-flavor transformations generates only  $\mathbb CP^{1}$ orbits (apart from the case where only $\gamma^{8}$ is non-zero, which is invariant).

We stress again that the cases listed above are just a consequence of boundary conditions when we search for closed loops at spatial infinity. Moreover, all the cases are topologically equivalent. In the next Section, we will study how the cases above translate into stable vortex configurations.

\subsubsection{BDM vortex}

The ``BDM case'' corresponds to the vortex studied by Balachandran, Digal and Matsuura in Ref.~\cite{Balachandran:2005ev}. Since we only need $T^{8}$ to generate the correct winding for this vortex, the idea is to restrict the action~(\ref{eq:lagrangianEM}) to include only the gauge fields $A^{8}$ and $A^{em}$, by keeping all the other gauge fields to zero. Formally the action reduces to that of a $U(1)\times U(1)$ gauge theory, which we can then express in terms of the massless and massive combinations~(\ref{eq:diagon}):
\begin{eqnarray}
SU(3)_{C}\times U(1)_{\textsc{em}}\rightarrow U(1)_{8}\times U(1)_{\textsc{em}}\simeq U(1)_{0}\times U(1)_{M}\,,
\end{eqnarray}
where the arrow means that we truncate the model to its Abelian sub algebra.   
The Lagrangian reads
\begin{align}
\mathcal L &=  \Tr\left[-\frac14\frac32 F^{0}_{ij}F^{0ij}\right] 	+\nonumber \\	
&- \Tr\left[\frac14\frac32 F^{M}_{ij}F^{Mij}+ K_{1}\nabla_{i} \Phi^{\dagger}\nabla^{i} \Phi -\lambda_{2}(\Phi^{\dagger}\Phi)^{2}+m^{2}\Phi^{\dagger}\Phi  \right] -\lambda_{1}(\Tr[\Phi^{\dagger}\Phi])^{2}-\frac{3m^{4}}{4(3\lambda_{1}+\lambda_{2})},   \label{eq:diaglagBDM}
\end{align}
with
\begin{align}
 \nabla_{i}=\p_{i}-ig_{M}A^{M}T^{M},
\quad g_{M} \equiv \sqrt{e^{2}+3 g_{s}^{2}/2},\quad T^{M}\equiv T^{EM}\,,
\end{align}
which is applicable only for diagonal configurations for the order parameter $\Phi$. In the action above, the massless field decouple completely, as expected, while the massive field $A_{M}$ couples to $\Phi$ in the standard way. The BDM vortex is constructed analogously to~(\ref{eq:antiredvortex}) with the only difference from the un-coupled case being the new coupling constant $g_{M}$ instead of $g_{s}$:
\begin{align}
& \Phi(r,\varphi)_{\textsc{bdm}}= \Delta_{\textsc{cfl}}
\begin{pmatrix}
 e^{i\varphi}f(r) &  0 & 0  \\
0  &  g(r) &  0 \\
 0 & 0  & g(r)   
\end{pmatrix} , \nonumber \\ 
& A_{i}^{M}T^{M}= \frac1{g_{M}}\frac{\epsilon_{ij} x^{j}}{r^{2}}[1-h(r)] \, T^{M}
, \quad A_{i}^{0}=0 \,.
\label{eq:BDMvortex}
\end{align}
 As mentioned in Section 2, the tension of the vortex decreases monotonically with the gauge coupling. Since we have $g_{M}>g_{s}$, the BDM vortex has a smaller tension as compared to the corresponding vortex in the un-coupled case. 
\subsubsection{$\mathbb CP^{1}$ vortex}

The $\mathbb CP^{1}$ case is considered for the first time in this paper. We have identified a vortex configuration which is a solution of the equations of motion that corresponds to these new boundary conditions. To see this, we notice that, similarly to the previous case, we can consistently restrict the action to include only the gauge fields corresponding to generators commuting with $T^{8}$. This corresponds to formally reduce to the case
\begin{align}
SU(3)_{C}\times U(1)_{\textsc{em}}\rightarrow SU(2)_{C}\times U(1)_{8}\times U(1)_{\textsc{em}}\simeq SU(2)_{C}\times U(1)_{0}\times U(1)_{M}\,,
\end{align}
similarly as what we have done for the BDM case. The truncated lagrangian now is the following:
\begin{align}
	\mathcal L &= \Tr\left[-\frac14\frac32 F^{0}_{ij}F^{0ij}\right]+\nonumber \\	
	&- \Tr\left[\frac14 \frac32 F^{M}_{ij}F^{Mij} -\frac14 F^{b}_{ij}F^{bij} 	+K_{1} \nabla_{i} \Phi^{\dagger}\nabla^{i} \Phi -\lambda_{2}(\Phi^{\dagger}\Phi)^{2}+m^{2}\Phi^{\dagger}\Phi  \right] -\lambda_{1}(\Tr[\Phi^{\dagger}\Phi])^{2}+ \nonumber \\
	&-\frac{3m^{4}}{4(3\lambda_{1}+\lambda_{2})}\,;\nonumber \\
	& \nonumber \\
	& \nabla_{i}=\p_{i}-ig_{M}A^{M}T^{M}-i g_{s}A^{b}T^{b} \, , \quad b=1,2,3 \, ; \quad [T^{b},T^{8}]=0 \,,
  	\label{eq:diaglagCP1}
\end{align}
where the index $b$ is relative to the $SU(2)_C$ factor.
As before, the massless combination decouples completely, and we can set it to zero. 
The simplest vortex configuration of the type considered here 
has the following diagonal form:
\begin{align}
	& \Phi(r,\varphi)_{\mathbb CP^{1}_{+}}= \Delta_{\textsc{cfl}}
\left(
\begin{array}{ccc}
 g_{1}(r)  &  0 & 0  \\
0  &  e^{i\varphi}f(r)  &  0 \\
 0 & 0  & g_{2}(r)   
\end{array}
\right) \, , \nonumber \\ 
	&  A_{i}^{M}T^{M}=-\frac12\frac1{g_{M}}\frac{\epsilon_{ij} x^{j}}{r^{2}}[1-h(r)] T^{M}, \nonumber \\
	& A_{i}^{3}T^{3}=\frac{1}{\sqrt2}\frac1g_{s}\frac{\epsilon_{ij} x^{j}}{r^{2}}[1-l(r)]\,T^{3}\,. \label{eq:CP1ansatz}
\end{align}
As explained in Eq.~(\ref{eq:CFModuli}), we can generate a full $\mathbb CP^{1}$ of solutions by applying $SU(2)_{C+F}$ rotations to the configuration above. Once the ansatz \eqref{eq:CP1ansatz} is inserted into the equations of motion~(\ref{eq:motion}), we get the equations written in the Appendix. We have numerically solved these equations, and determined the energy of the vortex configuration. The tension, as schematically shown in Fig.~\ref{fig:ToricTens}, is higher for the $\mathbb CP^{1}$ vortices than for the BDM vortex. This result can be intuitively understood if we recall the observation of the previous Section, where we noticed that the tension of a color vortex decreases monotonically with the gauge coupling. The $\mathbb CP^{1}$ vortex is built with both $g_{s}$ and $g_{M}$ couplings, thus, the interactions depending on $g_{s}<g_{M}$ contribute to increase the tension with respect to a vortex built exclusively from $g_{M}$.

Since the $\mathbb CP^{1}$ vortex is a solution of the equations of motion, it is a critical configuration for the energy density functional. In Section~\ref{sec:stability}, we will argue that it must be a local minimum, and thus a metastable configuration.

\subsubsection{Pure color vortex}

\begin{figure}[htbp]
\begin{center}
\includegraphics{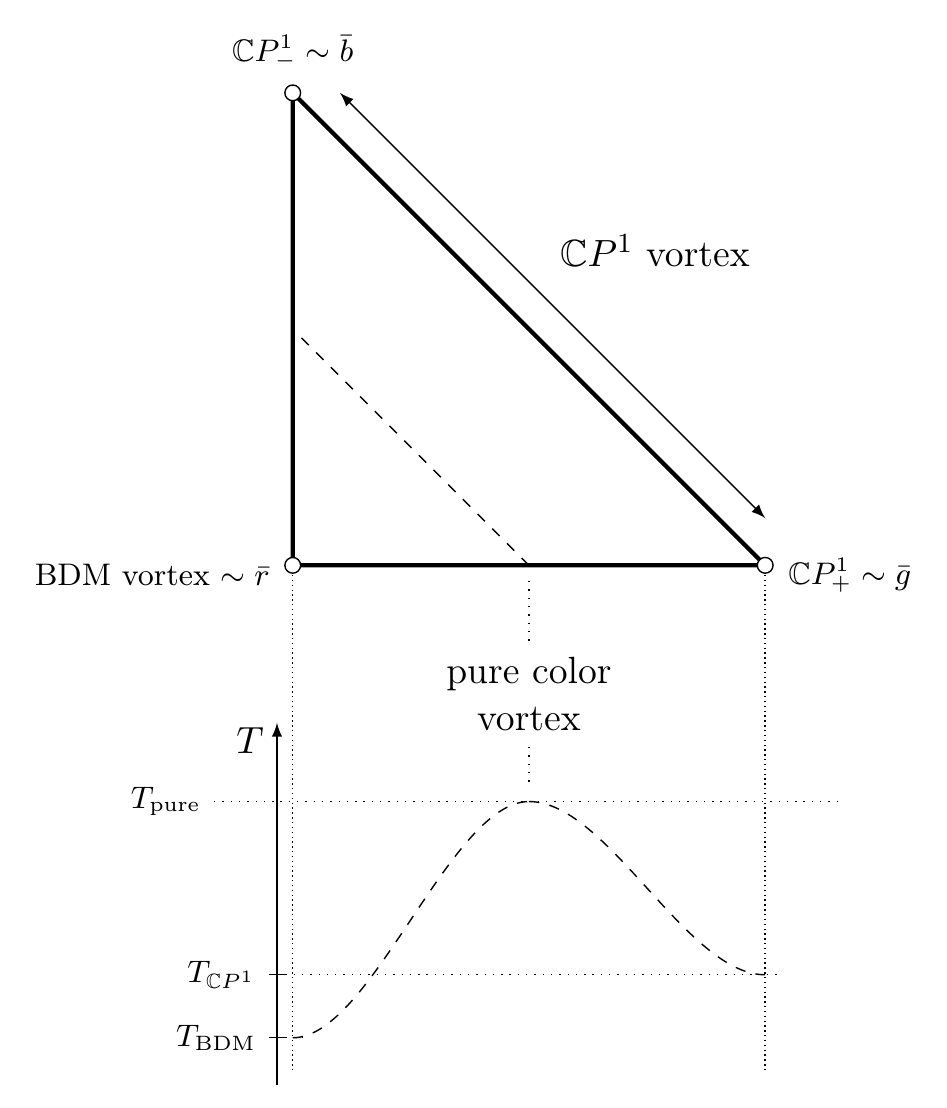}
	\caption{The three different special vortex solution explained in the text. The BDM vortex sits in the bottom left corner of the toric diagram and has the lowest tension. The other two diagonal vortices are located at the remaining corners and are connected by color-flavor transformations. All the upper right edge of the diagram represents the whole family of vortex configurations related by this $SU(2)_{C+F}$ transformations, namely a $\mathbb{C}P^{1}$ moduli space. This vortices have a slightly larger tension compared to the BDM configuration. The dashed line in the middle of the diagram corresponds to vortices not winding along $T^{8}$ direction in color group. These vortices have the largest tension and do not involve electromagnetic gauge field. We reported a qualitative behavior of the potential between these solutions under the diagram.}
	\label{fig:ToricTens}
\end{center}
\end{figure}

Let us now come to the third case, the ``pure color'' case. This situation is realized when the boundary conditions are the same as the case without electromagnetic coupling. This means that at infinity, only the non-Abelian gauge fields gives non-trivial winding, and thus non-vanishing fluxes, while the electromagnetic gauge field is zero everywhere. However, this situation is generically not compatible with the coupled equations of motion. An easy way to see this is the fact that boundary conditions in the pure color case imply a non-zero flux for the massless field $A_{i}^{0}$. Since $A_{i}^{0}$ is massless and unbroken, 
there is no topology, compatible with the equations of motion, stabilizing and confining the flux.
We thus expect that no solutions exist in general for the pure case. However, a special configuration that we call ``pure color vortex'' is an exception to this statement. This configuration is defined as the one corresponding to the following boundary conditions:
\begin{eqnarray}
&e^{i \gamma^{a}(2\pi)T^{a}}=e^{-i\theta(2\pi)}\,; \quad \gamma^{8}\equiv 0\,. &
\end{eqnarray}
Notice that the conditions above are perfectly consistent. We can take any configuration of $\gamma^{a}$ phases and set $\gamma^{8}$ to zero using a gauge(-flavor) transformation. This means that we can consistently set:
\begin{eqnarray}
A_{i}^{8}=A_{i}^{em}=0, \quad \Rightarrow \quad A_{i}^{M}=A_{i}^{0}=0\,.
\end{eqnarray}
In fact, $A_{i}^{em}=0$ means that for the pure color vortex we can consistently restrict the action 
by simply dropping all the terms involving $A_{i}^{em}$. As a consequence, the pure color vortex is exactly the same configuration as we get in the un-coupled case. Moreover, it satisfies the full equations of motion of the coupled case, because of the consistent restriction. Notice that the dependence of the electromagnetic gauge coupling disappears also completely from the restricted action. This means that the tension of the pure color vortex is also the same as the tension of the un-coupled vortices, involving only $g_{s}$. As represented in Fig.~\ref{fig:ToricTens}, the tension of this vortex is larger than both the BDM and the $\mathbb CP^{1}$ vortices.

Because of this very fact we are lead to conclude that the pure vortex is in fact a stationary point for the energy functional, but it corresponds to a local maximum. Both the BDM and the $\mathbb CP^{1}$ vortices are on the other hand local minima, with the BDM vortex being the absolute minimum. The $\mathbb CP^{1}$ vortices are however metastable, and if long-lived can play a crucial role to the physics  of the CFL superconducting phase together with the BDM vortex. The whole situation is schematically summarized in Fig.~\ref{fig:ToricTens}. Numerical values of the tensions are compared as 
\begin{eqnarray}
&\mathcal T_{\text{pure}}-\mathcal T_{\textsc{bdm}}=0.0176 \,{\rm MeV}^{2}, \quad 
 \mathcal T_{\text{pure}}- \mathcal T_{\mathbb CP^{1}}= 0.0044 \,{\rm MeV}^{2} \, ,
 \label{eq:tensions}
\end{eqnarray}
for the same ``realistic'' choice of parameters made in Fig.~\ref{fig:tension}, where in addition we have chosen $m_{G}=92$ MeV and a value for the electromagnetic coupling constant of $e^{2}=1/137$. Notice that the expressions shown above do not depend on the infrared regulator $L$. The diverging parts of the tensions, in fact, are equal and cancel out in the differences.

We conclude this Section by recalling that the low energy physics of an uncoupled color vortex is described by a  $\mathbb CP^{2}$ non-linear sigma model, as shown in Ref.~\cite{Eto:2009bh}. When we consider the effects of the electromagnetic coupling, the effective theory should become a $U(1)$ gauged $\mathbb CP^{2}$ model~\cite{Hirono:2012ki}, where we have to include also the effective potential sketched in Fig.~\ref{fig:ToricTens}.

\subsection{Magnetic fluxes}

There are two main differences between color magnetic flux tubes with and without electromagnetic coupling. The first one, as we have already examined, is the lifting of the moduli space $\mathbb CP^{2}$ to leave  the stable BDM vortex and the family of  metastable degenerated $\mathbb CP^{1}$ vortices. The second is the fact that coupled vortices now carry a non-trivial electromagnetic flux. This is given by the fact that coupled vortices are made of the massive field $A_{i}^{M}$, which is in turn a linear combination of color and electromagnetic gauge fields. 

The BDM vortex carries a quantized $A_{i}^{M}$ flux:
\begin{eqnarray}
& \displaystyle \Phi^{M}_{\textsc{bdm}}= \oint \vec A^{M}\cdot  d\vec l=\frac{2 \pi}{g_{M}}\,.
 \end{eqnarray}
Because of the mixing in the ground state, this means the following non-quantized fluxes for the color and electromagnetic fields:
\begin{align}
& \Phi^{8}_{\textsc{bdm}}=\sqrt\frac{2}{3}\frac{1}{1+\delta^{2}}\frac{2 \pi}{g_{s}},\quad \Phi^{\textsc{em}}_{\textsc{bdm}}=\frac{\delta^{2}}{1+\delta^{2}}\frac{2 \pi}{e}\, , \quad \delta^{2} \equiv \frac23 \frac{e^{2}}{g_{s}^{2}} \, .
 \end{align}
 The fluxes of the $\mathbb CP^{1}$ vortices can be similarly determined:
\begin{align}
&\Phi^{M}_{\mathbb CP^{1}}=-\frac12 \Phi^{M}_{\textsc{bdm}} \, 
 \quad \Rightarrow \quad   
\Phi^{8}_{\mathbb CP^{1}}=-\frac12 \Phi^{8}_{\textsc{bdm}},\quad 
\Phi^{\textsc{em}}_{\mathbb CP^{1}}=-\frac12 \Phi^{\textsc{em}}_{\textsc{bdm}}, \nonumber \\
 &\Phi^{3}_{\mathbb CP^{1}_{+}} \, =  \,  \oint \vec A^{3}\cdot  d\vec l=\frac{1}{\sqrt2}\frac{2 \pi}{g_{s}}\,,\quad \Phi^{3}_{\mathbb CP^{1}_{-}}=-\Phi^{3}_{\mathbb CP^{1}_{+}}\,.
 \end{align}
 Moreover, the quantized circulations of the BDM  and $\mathbb CP^{1}$ vortices are the same as that of usual un-coupled vortices, and equal to $c_{B}/3$.

Notice that the expressions determined above imply that a color-neutral bound state of vortices necessarily carries also no electromagnetic flux, and vice versa. In particular, in the un-coupled case we need at least a bound state of three vortices, ($\bar r, \bar g, \bar b$) to obtain a color-less state, which is nothing but the $U(1)_{\rm B}$ vortex. In the coupled case, this ``minimal'' color-less bound state is obtained with the combination ($BDM,\, \mathbb CP^{1}_{+},\,\mathbb CP^{1}_{-}$). The bound state carries an integer circulation $c_{B}=1$.

\begin{figure}[htbp]
\begin{center}
\includegraphics{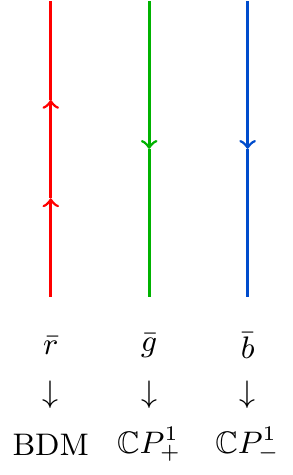}
	\caption{The three different types of vortices we consider. The BDM vortex is labelled as anti-red $\bar{r}$ vortex as explained in the text, and carries twice electromagnetic flux compared to the other solutions, the anti-blu $\bar{b}$ and anti-green $\bar{g}$ vortices.}
	\label{fig:VortColors}
\end{center}
\end{figure}


\subsection{Quantum mechanical decay}
As we have seen, the ${\mathbb C}P^1$ vortices are classically metastable because of the potential barrier 
by the presence of the pure color vortex. 
However here we show that the ${\mathbb C}P^1$ vortices are quantum mechanically unstable, and they can decay to the BDM vortex by quantum tunneling. 
We estimate this decay probability.

As discussed in the previous subsections, we do not expect static solutions of the equations of motion apart from the BDM, the $\mathbb CP^{1}$ and the pure color vortices. However, we can try to define an effective potential interpolating between the three types of solutions. At generic orientations, boundary conditions are the same as those for the un-coupled vortices. In fact, exactly the un-coupled vortex evaluated on the coupled action Eq.~(\ref{eq:lagrangianEM}) gives the same tension as that of un-coupled vortices. This value of the tension is an upper bound for vortex configurations that have a fixed boundary condition corresponding to a generic point in $\mathbb CP^{2}$. The tension of the configuration that really minimizes the energy, for that fixed boundary conditions, defines an ``effective'' potential on the  $\mathbb{C}P^{2}$ moduli space induced by the electromagnetic interactions. Moreover, since the tension of vortices is mainly modified by the contribution of the mixed $A_{i}^{M}$, which couples with a larger gauge coupling, we expect the qualitative behavior of the effective potential to be of the form represented in Fig.~\ref{fig:ToricTens}

This qualitative picture is enough to address some important features of coupled color-magnetic flux tubes. The fact that the potential has more than one local minimum allows for the existence of kinks interpolating between the two vortices corresponding to the various (meta)stable configurations; these kinks are interpreted as confined monopoles.  Moreover, the presence of kinks, and the higher tension of  $\mathbb CP^{1}$ vortices  with respect to the BDM vortex, implies a decay rate of the former vortices into the latter through quantum tunneling. This tunneling proceed by enucleation of kink/anti-kinks pairs along the vortex \cite{Preskill:1992ck,Shifman:2002yi,Voloshin:1985id}  and it is similar to the quantum decay of a false vacuum in 1+1 dimensions \cite{Voloshin:1985id}.  An analogous situation arises  for pure color vortices \cite{Eto:2011mk,Gorsky:2011hd}, where  the potential is generated by quantum non-perturbative effects of the  $\mathbb CP^{2}$ non-linear sigma model.

 The decay rate can be roughly estimated in our case by following the arguments of Refs.~\cite{Preskill:1992ck,Voloshin:1985id}. The enucleation of a couple of kinks costs an energy of order $M_{kink}$. Moreover, they are created at a critical distance $L_{crit}$ such that the energy cost for the pair production is balanced by the energy gain due to the presence of an intermediate vortex with smaller tension: $L_{crit} \Delta \mathcal T \sim M_{kink}$. The decay probability rate per unit length is thus:
\begin{equation}
P\sim e^{-M^{2}_{kink}/ \Delta \mathcal T}\,.
\end{equation}

We now apply the formula above to our specific case. The mass of the kinks can be estimated as being of the order of the square root of the height of the potential times the ``size'' $\beta$ of the moduli space: 
\begin{equation}
M_{kink}\sim \beta \sqrt{\mathcal T_{\text{pure}}-\mathcal T_{\mathbb CP^{1}}}\,.
\end{equation}
The quantity $\beta$ has been evaluated analytically and numerically in  Refs.~\cite{Eto:2011mk,Gorsky:2011hd}:
\begin{equation}
\beta \sim K_{1}^{2}/\lambda_{1}\sim \mu^{2}/T_{c}^{2},
\end{equation}
and it turns out to be large for our ``realistic'' regime, $\beta\sim2500$. The tension difference is given by $\Delta \mathcal T=\mathcal T_{\text{pure}}-\mathcal T_{\textsc{bdm}}$. We have already reported the numerical estimates of the  quantities above in Eq.~(\ref{eq:tensions}), for a special value of the couplings.
The decay probability is then:
\begin{equation}
P\sim e^{-\beta\,R},\quad R\equiv\frac{\mathcal T_{\text{pure}}-\mathcal T_{\mathbb CP^{1}}}{\mathcal T_{\text{pure}}-\mathcal T_{\textsc{bdm}}}
\label{eq:Ratio}
\end{equation}

Substituting the numerical values of Eq.~(\ref{eq:tensions}) we obtain $R\sim0.25$. We have also studied the dependence of this decay probability in terms of a more general set of values of the gauge couplings. The numerical results are shown in Fig.~\ref{fig:Ratio}. In the left panel  we have plotted the quantity $R$ as a function of the mass $m_{g}$, which depends on the gauge coupling $g_{s}$, where we have set $e^{2}=1/137$.  On the right panel, we show the same quantity as a function of the electromagnetic coupling $e$, where we have set $m_{g}=92$ MeV. We see that $R$ has a very mild dependence on the value of both $m_{g}$ and $e$. As shown in Fig.~\ref{fig:Ratio}, $R$ is always a quantity of order 1.  We have limited ourselves to consider large values of the gauge bosons ($m_{g}\gtrsim 10$ MeV) and   small values of the electromagnetic gauge coupling ($e\lesssim1$), as expected in realistic settings in the CFL phase. We thus see that for the range of values of the couplings considered, the ratio $R$ is never small enough to compensate the large ``moduli space'' factor $\beta$. We thus estimate the probability of decay of  $\mathbb CP^{1}$ vortices in BDM vortices to be exponentially small in realistic settings. 
\begin{figure}[htbp]
\begin{center}
\begin{tabular}{cc}
\includegraphics[height=4.5cm]{Ratio.pdf} & \includegraphics[height=4.5cm]{RatioCoup.pdf}
\end{tabular}
	\caption{The numerical value of the ratio $R$ of Eq.~(\ref{eq:Ratio}). In the left panel, we have plot it as function of the mass $m_{g}$, fixing the value of the electromagnetic gauge coupling to its realistic value $e^{2}=1/137$. In the right panel, we have plot the same quantity as a function of $e$, while we have fixed the value of the mass $m_{g}$ to the typical value: $m_{g}=92$ MeV. Notice that the value of the ratio $R$ is always of order 1 for the wide range of the physical parameter values chosen.}
	\label{fig:Ratio}
	\end{center}
\end{figure}

 Notice that the estimate made above is fully justified only in the case the size of the kink is negligible as compared to the critical length $L_{crit}$, while  in our case the two sizes are comparable. This estimate is very rough, and has to be corrected including possible Coulomb interactions between the kink/anti-kink pair. However, it should correctly capture the order of magnitude of the decay probability. 
It is an open problem to give a more precise estimate of the decay probability  of  $\mathbb CP^{1}$ vortices in the CFL with electromagnetic coupling and vanishing quark masses. However, as we shall see in the next Section, the effects of a non-vanishing strange quark mass overshadow the effects of the potential induced by electromagnetic interactions in more realistic settings. 


\section{The Comparison with the other Potential Terms
}
\label{sec:stability}

In this Section, we compare the potential generated semi-classically by the electromagnetic interactions with the other potentials, {\it i.e.}, 
the quantum mechanically induced potential and 
the potential induced by the strange quark mass.

\subsection{Quantum mechanical potential}
 Our task here is to compare the qualitative potential generated semi-classically by the electromagnetic interactions to the quantum potential generated by $\mathbb CP^{2}$. The quantum potential has three vacua given by Ref.~\cite{Witten:1978bc,Witten:1998uka}
\begin{equation}
E_{\mathbb CP^{2}}(k)\sim 3 \Lambda^{2}_{\mathbb CP^{2}}\left[1+const  \left( \frac{2 \pi k }{3} \right)^{2} \right],\quad k=0,1,2 \,,
\end{equation}
where $\theta$ is a quantum generated $\theta$ angle. The $\Lambda$ scale of the sigma model is given by 
\begin{equation}
\Lambda_{\mathbb CP^{2}}= \frac{m}{\sqrt{K_{1}}} \exp \left(  - c\, \frac{4 \pi}{3}\beta\right)= \frac{m}{\sqrt{K_{1}}} \exp \left(  - c\, \frac{4 \pi}{3}\frac{K_{1}^{2}}{\lambda_{2}}\right)\,.
\end{equation}
The constant $c$ has been evaluated numerically in Refs.~\cite{Eto:2011mk,Gorsky:2011hd} and it is found to be of order 1.  The vacua above are considered as coming from a quantum potential that oscillates between the minima above and barriers of potential of height $\Lambda_{\mathbb CP^{2}}$. This implies the existence of kinks of mass:
\begin{equation}
M_{kink}\sim \Lambda_{\mathbb CP^{2}}\,.
\end{equation}
In the formula above the large factor $\beta$ introduced in the previous Section appears again.  The numerical value of $\Lambda_{\mathbb CP^{2}}$ is thus exponentially small and negligible as compared to the semiclassical potential induced by electromagnetic interactions.

\subsection{Strange quark mass effects}\label{sec:StrangeMass}

So far we have considered the CFL phase at very high densities, where quark masses can be neglected, including the mass of the strange quark. The effect of a non-zero mass for the strange quark corresponds to the introduction of the following effective potential term in the Landau-Ginzburg description Eq.~(\ref{eq:lagrangianEM})~\cite{Iida:2003cc}:
\begin{eqnarray}
V_{m_{s}}=\epsilon \,\Tr\left[ \Phi^{\dagger} \left(\frac23 {\bf 1}_{2}+ T^{3} \right)\Phi  \right],\quad \epsilon=\frac{m_{s}^{2}}{2\pi^{2}}\log \frac{\mu}{T_{c}}\,.
\end{eqnarray}

The effects of this potential on the $\mathbb CP^{2}$ moduli space of pure color vortices have been studied in Ref.~\cite{Eto:2009tr}. It can be considered as generating the following effective potential
\begin{eqnarray}
V_{\mathbb CP^{2}}=\epsilon \int dx^{2}\,\Tr\left[\Phi^{\dagger}  T^{3} \Phi  \right]\equiv D (|\phi_{3}|^{2}-|\phi_{1}|^{2})\,.
\end{eqnarray}
Where $D$ is determined in terms of the vortex profile functions:
\begin{eqnarray}
D=\pi \epsilon \Delta_{\epsilon} \int_{0}^{\infty}dr \,r\,(f^{2}-g^{2})\,,\quad \Delta_{\epsilon}=\Delta_{\textsc{cfl}}(m^{2}\rightarrow m^{2}+2\epsilon/3)\,,
\end{eqnarray}
and $|\phi_{1}|^{2}+|\phi_{2}|^{2}+|\phi_{3}|^{3}=1$ parameterize the $\mathbb CP^{1}$s which, in the toric representation of $\mathbb CP^{2}$ are all parallel to the long side of the triangle. In our notations, $\bar r=(1,0,0)$, $\bar g=(0,1,0)$ and $\bar b=(0,0,1)$. The vortex with less energy turns out to be then the one with $|\phi_{2}|=1$, or the $\bar g=(0,1,0)$ vortex. When we turn on the electromagnetic interactions, this vortex corresponds to what we called the $\mathbb CP^{1}_{+}$ vortex. We then see that the introduction of strange quark mass makes the BDM vortex unstable through decay to the $\mathbb CP^{1}_{+}$ vortex. Our new solution is thus the one that can play the main role in the CFL phase at intermediate densities, when the mass of the strange quark cannot be neglected. Moreover, it removes the degeneration between the $\mathbb CP^{1}$ vortices.  We have compared the two effects with numerical simulations, and the results are:
\begin{eqnarray}
&\mathcal T=5897MeV^{2} ,\quad 
D = 4231MeV^{2} , \quad 
\Delta V = 0.0275MeV , 
\end{eqnarray}
with the usual choice of parameters and in addition $ m_{s} = 150MeV$.

We see that the potential induced by electromagnetic interactions is comparable with the effects of the strange quark mass only when the strange quark mass is very negligible as compared to the chemical potential ($\epsilon\ll m^{2}$). These densities are not realistic in neutron stars core, for example. We then have found that the  $\mathbb CP^{1}_{+}$, with its different flux with respect to the previously studied BDM vortex, is the the most relevant to the physics of neutron stars in their inner core.

\section{Conclusions and Discussion}\label{sec:Conclusions}
The non-Abelian vortex in the CFL phase has the degeneracy of the Nambu-Goldstone modes ${\mathbb C}P^2$, associated with the color-flavor locked symmetry 
$SU(3)_{\rm C+F}$ spontaneously broken down to a subgroup in the vicinity of the vortex, when the electromagnetic coupling is neglected.
Once the electromagnetic coupling is taken into account, 
the flavor symmetry $SU(3)_{\rm F}$ is explicitly broken. 
We have studied the electromagnetic coupling of the non-Abelian vortices and have found that the degeneracy of the ${\mathbb C}P^2$ modes is removed with the effective potential induced. 
We have found that the stable BDM vortex solution and a family of metastable vortex solutions parametrized by ${\mathbb C}P^1$, both of which carry ordinary magnetic fluxes in addition to color magnetic fluxes. We have discussed quantum mechanical decay of the metastable ${\mathbb C}P^1$ vortices by quantum tunneling. 
We also have compared the effective potential which we found with the other known potentials: the quantum mechanically induced potential 
and the potential induced by the strange quark mass.
We have found that it is more dominant than the former but is negligible compared with the latter. 
Therefore all vortices including the BDM vortex decay into one of ${\mathbb C}P^1$ vortices. 

The results we have found are important for the study of the physics of neutron stars. The very high rotation speed of these compact stars causes the formation of the non-Abelian semi-superfluid vortices we analyzed. These solitons arrange in a lattice and they are constrained to be together to cancel the total color flux. Each vortex carries also a non-zero magnetic flux due to electromagnetic coupling, but as summarized in Fig.~\ref{fig:VortColors} the total flux of this vortex bound state is zero. Then we can say that no magnetic field generated by vortex formation comes out from the CFL core. However, it is well known that neutron stars possess a huge magnetic field, which should penetrate in the core as well. Thus there should be some non-trivial interaction between the external magnetic field spreading into the core and the semi-superfluid vortices present there and it is thus an interesting problem to know how the system behaves in this kind of situation. 

Problems similar to those considered in this paper can be analyzed in supersymmetric models with non-Abelian gauge group~\cite{Auzzi:2003fs,Hanany:2003hp,Tong:2005un,Eto:2006pg,Shifman:2007ce,Tong:2008qd}
, where a global color-flavor locked symmetry is unbroken in the vacuum. Interesting physics can be found when gauging the whole or a part of the color-flavor group~\cite{Konishi:2012eq}.

\section*{Acknowledgments}

We would like to thank Minoru Eto, Yuji Hirono and Kenichi Konishi for a discussion.
This work of M.N. is supported in part by 
Grant-in Aid for Scientific Research (No.~23740198) 
and by the ``Topological Quantum Phenomena'' 
Grant-in Aid for Scientific Research 
on Innovative Areas (No.~23103515)  
from the Ministry of Education, Culture, Sports, Science and Technology 
(MEXT) of Japan. W.V. thanks the FTPI institute for theoretical physics for the very warm and friendly hospitality while this work was under preparation, and in particular thanks M. Shifman, A. Yung and A. Vainshtein for valuable discussions. M.N. thanks INFN, Pisa, for partial support and hospitality while this work was completed.

\appendix


\section{Appendix}

\subsubsection*{Phases and Fluxes}
A generic boundary condition for the order parameter, representing a closed loop in the global-gauge groups can be represented by the following phases
\begin{eqnarray}
&\left<\Phi(\infty,2\pi)\right> = e^{i \theta} e^{i \gamma^{a}T^{a}}\left<\Phi(\infty,0)\right>  \,  e^{i \alpha T^{\textsc{em}}}\,,  &
 \label{eq:phases}
\end{eqnarray}
which correspond to the following fluxes
\begin{eqnarray}
 &A_{i}^{a}\sim \frac{\gamma^{a}}{2\pi}\frac1g_{s}\frac{\epsilon_{ij} x^{j}}{r^{2}}T^{a},\quad A_{i}^{\textsc{em}}=\frac{\alpha}{2\pi}\frac1e\frac{\epsilon_{ij} x^{j}}{r^{2}}T^{\textsc{em}} &\nonumber \\
 &\Phi^{a}=2\frac{\gamma^{a}}{g_{s}},\quad  \Phi^{em}=2\frac{\alpha^{2}}{e}.\quad &
\end{eqnarray}

\subsubsection*{Equations of motion for the un-gauged and BDM vortices}

With the usual ansatz
\begin{equation}
\Phi(r,\varphi)_{\bar r}= \Delta_{\textsc{cfl}}
\left(
\begin{array}{ccc}
 e^{i\varphi}f(r) &  0 & 0  \\
0  &  g(r) &  0 \\
 0 & 0  & g(r)   
\end{array}
\right),\quad A_{i}^{8}T^{8}=\sqrt\frac23 \frac1g_{s}\frac{\epsilon_{ij} x^{j}}{r^{2}}[1-h(r)] \, T^{8} ,
\end{equation}
we obtain the usual equations for the profile functions $f(r)$, $g(r)$ and $h(r)$:
\begin{eqnarray}
&f''+\frac{f'}{r}-\frac{(2h+1 )^{2}}{9 r^{2}}f-\frac{m_{\phi}^{2}}{6}f(f^{2}+2 g^{2}-3)-\frac{m_{\chi}^{2}}{3}f(f^{2}-g^{2})=0; \nonumber \\
&g''+\frac{g'}{r}-\frac{(h-1 )^{2}}{9 r^{2}}g-\frac{m_{\phi}^{2}}{6}g(f^{2}+2 g^{2}-3)+\frac{m_{\chi}^{2}}{6}g(f^{2}-g^{2})=0; \nonumber \\
& h''-\frac{h'}{r}-\frac{m_{g}^{2}}{3}\left( g^{2}(h-1)+f^{2}(2h+1)  \right)\,,
\end{eqnarray}
with
\begin{eqnarray}
m^{2}_{g}=2 g^{2}_{s}\Delta_{\textsc{cfl}}^{2}K_{1}, \quad m^{2}_{\phi}=\frac{2 \mu^{2}}{K_{1}},\quad m_{\chi}^{2}=\frac{4 \lambda_{2}\Delta_{\textsc{cfl}}^{2}}{K_{1}},\quad m_{\varphi}^{2}=0\,.
\end{eqnarray}
The equations for the BDM vortex are obtained by just substituting $T^{8}\rightarrow \sqrt{2/3}T^{M}$ and $g_{s}\rightarrow g_{M}$.

\subsubsection*{BDM vortex in terms of $A_{i}^{8}$ and $A_{i}^{em}$}

The vortex with minimum energy, including the BDM vortex,  is constructed in a background where the massless gauge field $A_{i}^{0}$ vanishes. We then have for the corresponding phases:
\begin{eqnarray}
&-\sqrt{\frac{3}{2}}\frac{g_{s}}{e}\alpha +\frac{e}{g_{s}}\gamma^{8} =0; \quad
\gamma^{8}/\sqrt6+\alpha/3=2\pi/3;& \nonumber \\
\nonumber \\
&\gamma^{8}=\sqrt\frac23\frac{2\pi}{1+\delta^{2}} \quad \alpha=\delta^{2}\frac{2\pi}{1+\delta^{2}},\quad  \delta^{2}\equiv\frac23 \frac{e^{2}}{g_{s}^{2}}\,.&
\end{eqnarray}
Then we have 
\begin{eqnarray}
A_{i}^{8}=\sqrt\frac23\frac{1}{1+\delta^{2}}  \frac1g_{s}\frac{\epsilon_{ij} x^{j}}{r^{2}}[1-h(r)] \, T^{8}\,; \quad
A_{i}^{em}=\frac{\delta^{2}}{1+\delta^{2}} \frac1e\frac{\epsilon_{ij} x^{j}}{r^{2}}[1-l(r)]\,T^{\textsc{em}} .
\end{eqnarray}

\subsubsection*{Equations of motion for the $\mathbb CP^{1}$ vortices}
By substituting the following ansatz:
\begin{eqnarray}
&&\Phi(r,\varphi)_{\mathbb CP^{1}_{+}}= \Delta_{\textsc{cfl}}
\left(
\begin{array}{ccc}
 g_{1}(r)  &  0 & 0  \\
0  &  e^{i\varphi}f(r)  &  0 \\
 0 & 0  & g_{2}(r)   
\end{array}
\right),\nonumber \\ 
&&  A_{i}^{M}T^{M}=-\frac12\frac1{g_{M}}\frac{\epsilon_{ij} x^{j}}{r^{2}}[1-h(r)] T^{M} , \quad
 A_{i}^{3}T^{3}=\frac{1}{\sqrt2}\frac1g_{s}\frac{\epsilon_{ij} x^{j}}{r^{2}}[1-l(r)]\,T^{3}\,,
\end{eqnarray}
into the equations of motion~(\ref{eq:motion}) we obtain:
\begin{align}
&f''+\frac{f'}{r}-\frac{(1+h/2+3/2l )^{2}}{9 r^{2}}f-\frac{m_{\phi}^{2}}{6}f(f^{2}+ g_{1}^{2}+g_{2}^{2}-3)-\frac{m_{\chi}^{2}}{3}f(f^{2}-g_{1}^{2}/2-g_{2}^{2}/2)=0; \nonumber \\
&g_{1}''+\frac{g_{1}'}{r}-\frac{(h-1 )^{2}}{9 r^{2}}g_{1}-\frac{m_{\phi}^{2}}{6}g_{1}(f^{2}+g_{1}^{2}+g_{2}^{2}-3)-\frac{m_{\chi}^{2}}{3}g_{1}(g_{1}^{2}-f^{2}/2-g_{2}^{2}/2)=0; \nonumber \\
&g_{2}''+\frac{g_{2}'}{r}-\frac{(1+h/2-3/2l )^{2}}{9 r^{2}}g_{2}-\frac{m_{\phi}^{2}}{6}g_{2}(f^{2}+g_{1}^{2}+g_{2}^{2}-3)-\frac{m_{\chi}^{2}}{3}g_{2}(g_{2}^{2}-f^{2}/2-g_{1}^{2}/2)=0; \nonumber \\
& h''-\frac{h'}{r}-\frac{m_{M}^{2}}{3}\left( 2g_{1}^{2}(h-1)+f^{2}(1+h/2+3/2l )+g_{2}^{2}(1+h/2-3/2l )  \right)=0\,; \nonumber \\
& l''-\frac{l'}{r}-\frac{m_{3}^{2}}{3}\left( f^{2}(1+h/2+3/2l )-g_{2}^{2}(1+h/2-3/2l )  \right)=0\,. 
\end{align}

\newpage	

 \bibliography{Bibliographysmart-1}
\bibliographystyle{nb}

\end{document}